\documentclass[runningheads]{llncs} 
\usepackage[T1]{fontenc} 
\usepackage{graphicx} 
\usepackage{amsmath} 
\usepackage{float}
\begin{document}
\title{Empirical Research on Utilizing LLM-based Agents for Automated Bug Fixing via LangGraph 
}

\author{Jialin Wang\inst{1}  \and
Zhihua Duan\inst{2 }   }
 
\institute{
Executive Vice President,Ferret Relationship Intelligence\\Burlingame, CA 94010, USA \\
\email{jialinwangspace@gmail.com}\\
\url{https://www.linkedin.com/in/starspacenlp/} 
\and
Intelligent Cloud Network Monitoring Department \\
China Telecom Shanghai Company,Shanghai, China\\
\email{duanzh.sh@chinatelecom.cn}\\
}

\maketitle              
\begin{abstract}
This paper presents a novel framework for automated code generation and debugging, designed to improve accuracy, efficiency, and scalability in software development. The proposed system integrates three core components—LangGraph, GLM-4-Flash, and ChromaDB—within a four-step iterative workflow to deliver robust performance and seamless functionality.

LangGraph serves as a graph-based library for orchestrating tasks, providing precise control and execution while maintaining a unified state object for dynamic updates and consistency. It supports multi-agent, hierarchical, and sequential processes, making it highly adaptable to complex software engineering workflows. GLM-4-Flash, a large language model, leverages its advanced capabilities in natural language understanding, contextual reasoning, and multilingual support to generate accurate code snippets based on user prompts. ChromaDB acts as a vector database for semantic search and contextual memory storage, enabling the identification of patterns and the generation of context-aware bug fixes based on historical data.

The system operates through a structured four-step process: (1) Code Generation, which translates natural language descriptions into executable code; (2) Code Execution, which validates the code by identifying runtime errors and inconsistencies; (3) Code Repair, which iteratively refines buggy code using ChromaDB’s memory capabilities and LangGraph’s state tracking; and (4) Code Update, which ensures the code meets functional and performance requirements through iterative modifications.

By combining advanced task orchestration, semantic reasoning, and memory-driven insights, the system achieves precise and iterative debugging, significantly enhancing software development workflows. This work represents a substantial advancement in automated software engineering, addressing critical challenges in code reliability and runtime error resolution.

\end{abstract}
 
\section{Introduction}
In modern software development, code writing and debugging are crucial steps. With the rapid advancement of artificial intelligence technology, large language models are increasingly being applied in the fields of code writing and debugging. However, traditional methods of code generation and debugging often require extensive manual intervention, which not only reduces development efficiency but also increases the likelihood of errors. To address these challenges, we propose a code self-repair system based on LangGraph and large language models, which simplifies the development process by automatically debugging and fixing bugs. By integrating a BUG vector knowledge base, the system not only improves the accuracy of bug fixing but also ensures the safety of code modifications, precise state tracking, and reliable code verification.

The system covers multiple aspects, including code generation, code execution, code repair, and code updating. During the code generation phase, users or developers can describe the desired functionality in natural language. Based on the GLM-4-Flash large language model, the system interprets the semantics of the input natural language and generates corresponding code snippets. For example, Figure 1 illustrates the process of generating a simple division function code, but it does not handle the case where the divisor is zero, which could lead to a ZeroDivisionError at runtime.

\begin{figure}
\centering
\includegraphics[width=1\textwidth]{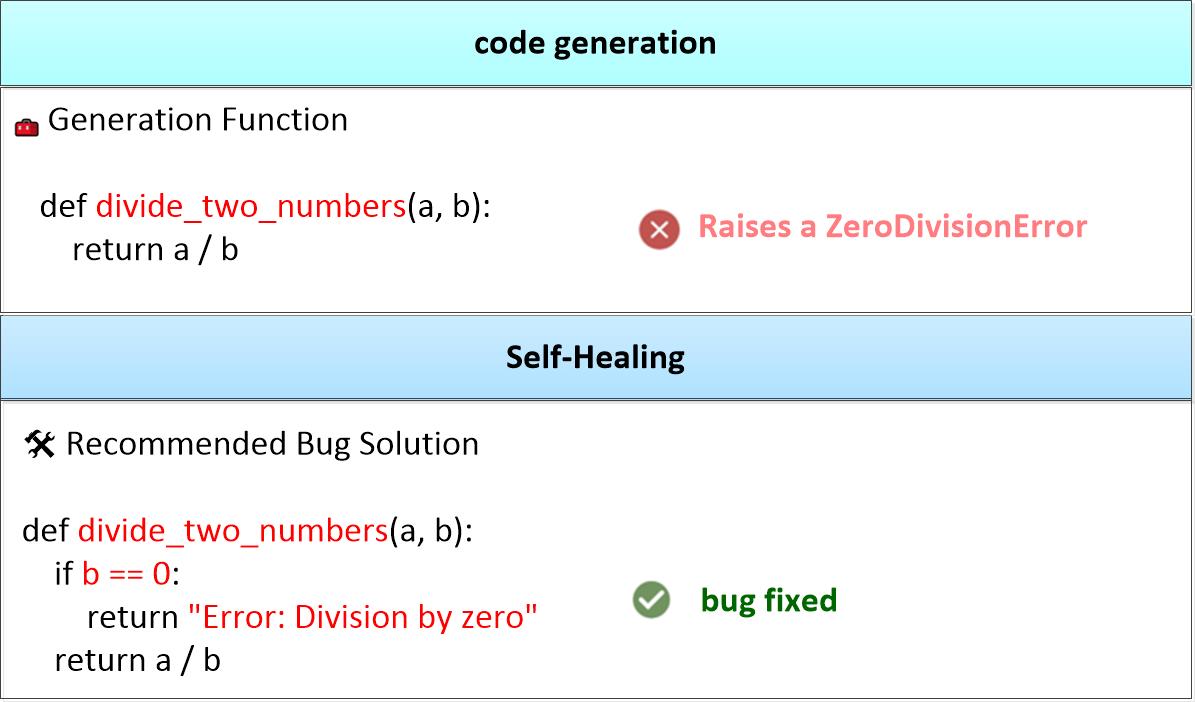}
\caption{Code Self-Repair Example Based on LangGraph.} \label{fig1}
\end{figure}

During code execution, if issues arise, the system collects error information and utilizes ChromaDB for memory search, filtering, creation, and updating operations to enhance the efficiency and accuracy of bug fixing. Then, the system feeds the code and error information back to the GLM-4-Flash model, which repairs the erroneous code and generates the corrected version. In Figure 1, the self-repair mechanism addresses the division-by-zero error by adding a conditional check; if a zero divisor is detected, it returns an error message instead of throwing an exception.

To validate the effectiveness of the system, we designed a series of experiments using GLM-4-Flash as the base model and implemented the entire process of code generation and bug fixing based on LangGraph. We also provided two examples that demonstrate the code generation and bug fixing processes for calculating the area of a triangle and dividing two numbers. This study not only offers a new automated solution for code generation and debugging but also opens up new avenues for the application of large language models in software development.

\section{Related Work}
\subsection{Software Engineering Evaluation}
The research team constructed SWE-bench, which includes 2,294 GitHub issues and corresponding pull requests from 12 popular Python repositories. The language model is required to edit the code repository based on the issue descriptions to resolve the problems\cite{SWE-bench}.
An empirical analysis was conducted on the SWE-bench dataset, revealing issues such as solution leakage and test cases within the dataset. Improved versions, SWE-bench Lite and SWE-Bench Verified, were proposed\cite{SWE-Bench2}.

\subsection{LLM based bug fixing system}
Systems based on large language models (LLM) can automatically modify code based on bug reports, utilizing the natural language processing capabilities of LLMs to achieve bug fixes.

MarsCode Agent is an error repair system developed by ByteDance. The system combines a code knowledge base, software analysis technology, and large language models (LLM). It uses LLM to generate candidate patches and selects the final patch from these candidates through scripts to achieve error repair\cite{Empirical}\cite{MarsCode}.Alibaba Lingma Agent is an error repair system developed by Alibaba. It constructs a knowledge graph for code repositories and uses LLM to perform Monte Carlo tree search based on problem information, locating code snippets related to the issue throughout the entire code repository\cite{Lingma}.The AutoCodeRover bug repair system uses spectrum-based fault localization (SBFL) for defect localization, provides LLM with features related to the problem, and then fixes code errors\cite{AutoCodeRover}.Agentless + RepoGraph is a defect repair system that combines code knowledge graphs with agentless methods\cite{RepoGraph}.

This study focuses on code defect repair and proposes a solution based on LangGraph and large language models. The solution utilizes LangGraph to execute code nodes, indexes and searches problem reports through a vector database, and facilitates an interactive bug repair process based on large models.

\section{ Systematic Method Design}
LangGraph is based on the tasks of large models, using graphs for precise control and generating applications through graph-based compilation. During the execution of tasks, it can maintain a unified state object that continuously updates the state as the business logic runs. LangGraph is a powerful and easy-to-use library for building stateful, multi-agent applications, making it highly suitable for constructing various types of AI applications, with advantages such as ease of use, flexibility, and scalability.

Figure 2 illustrates the code self-repair mechanism based on LangGraph and large models. This study utilizes LangGraph's automated bug resolution system to simplify the development process through automatic debugging by large models. By leveraging a BUG vector knowledge base, the precision of bug fixing is improved, while also ensuring the safe application of code modifications, precise state tracking, and reliable code verification. The system mainly consists of four major components: code generation, code execution, code repair, and code updating.
\begin{figure}
\centering
\includegraphics[width=1\textwidth]{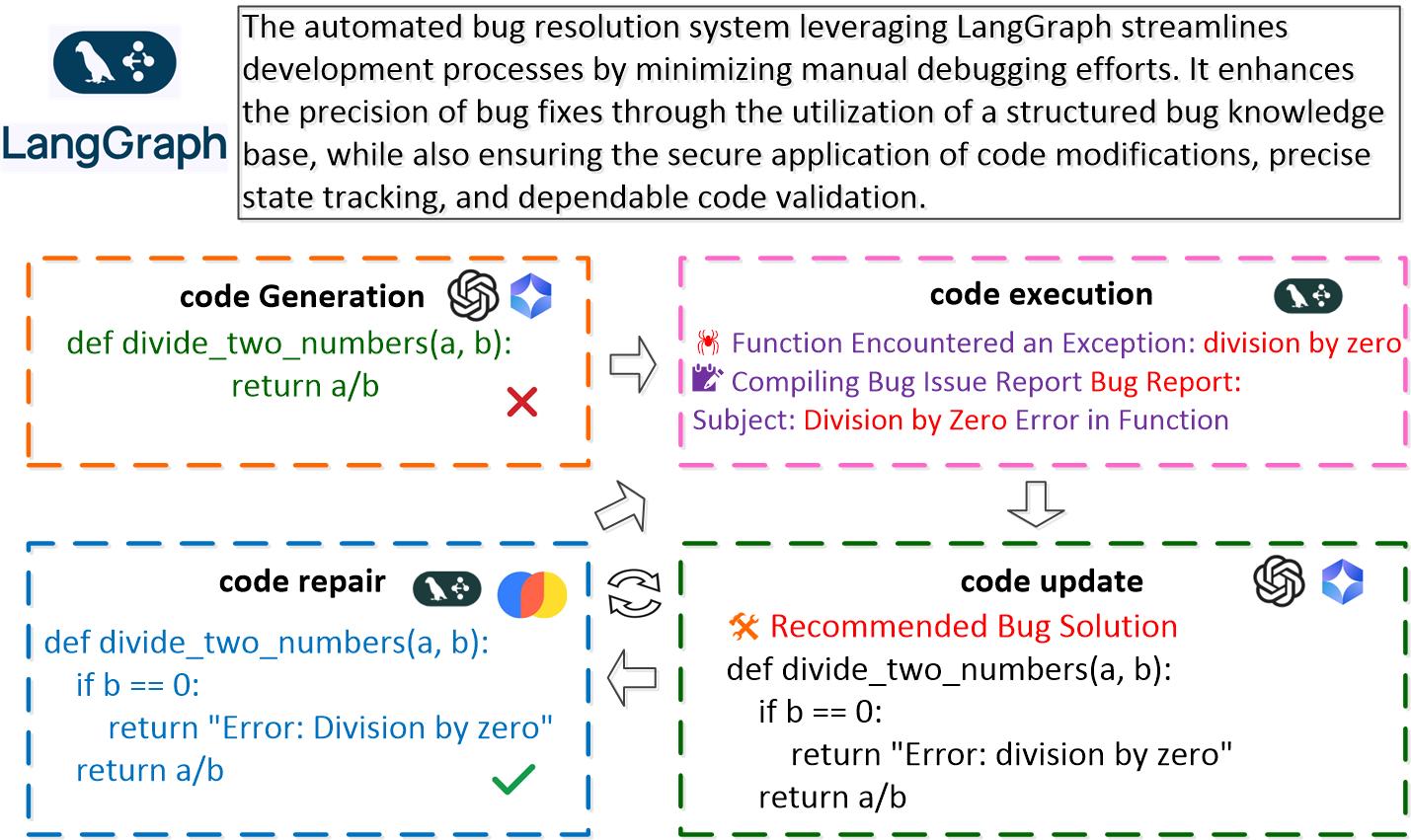}
\caption{Schematic diagram of code self-repair based on LangGraph and large models.} \label{fig1}
\end{figure}

\subsection{code generation}

Users or developers clearly describe the desired functionality and application scenarios in natural language, such as "Please implement a Python function for performing division between two numbers." This requirement text is submitted to LangGraph's code generation agent node. Based on the GLM-4-Flash large model, it can understand the semantics of the input natural language and generate the corresponding code snippet.

\subsection{code execution}
After generating the code, the code execution node in LangGraph is used for testing. LangGraph sets the function code status and function parameter status information in the global state information, and runs according to the logic set by the code. During this process, various output results and potential bug information are generated. For example, when performing a division operation between two numbers, if the divisor is 0, an error message indicating that the divisor cannot be 0 will be displayed. LangGraph records the bugs encountered during the execution process in a vector database, providing memory information for subsequent code optimization.

\subsection{code update}
When issues arise during code execution, such as syntax errors, runtime errors, or logic that does not meet expectations, the error messages and the program's abnormal behavior are collected by LangGraph and fed back to the GLM-4-Flash large model. Leveraging its powerful code capabilities and referring to correct programming patterns, the large model makes targeted modifications to the erroneous parts of the code, generating a corrected version. For example, in the code for dividing two numbers, if an exception is reported for a divisor of zero, the large model will add a conditional check for a divisor of zero to fix this bug. In this process, operations such as memory search, memory filtering, memory creation, and memory updating based on ChromaDB are also involved. The vector database is crucial for machine learning systems because it enables semantic search capabilities, supports efficient similarity calculations, has good scalability when handling large datasets, maintains contextual relationships, and aids in pattern recognition.

\subsection{code repair}
After LangGraph integrates the new code following bug fixes, it updates the global state of the function code in the code repair node, and sets the updated code based on this. The process then enters the execution phase again for verification, and this cycle continues until the code runs smoothly and essentially meets the original functional requirements, ensuring that the entire code system continues to meet the needs of practical applications.

\section{Experimental Design and Methods}
This experiment uses GLM-4-Flash as the base model. The GLM-4-Flash language model excels in web search, long context processing, and multilingual support, making it suitable for a variety of application scenarios such as intelligent Q\&A, summary generation, and text data processing.

The entire process of code generation and bug fixing is implemented based on LangGraph. LangGraph provides a highly customizable cognitive architecture that can adapt to various tasks and supports multiple control methods, including single-agent, multi-agent, hierarchical, and sequential execution, enabling robust handling of complex real-world scenarios.

Figure 3 shows a flowchart of the code generation and bug fixing process based on LangGraph. It describes an intelligent agent system for automatically correcting bugs based on LangGraph, GLM-4-Flash, and ChromaDB, achieving the full process of code generation, code execution, bug reporting, and bug fixing.

\begin{figure}
\centering
\includegraphics[width=0.6\textwidth]{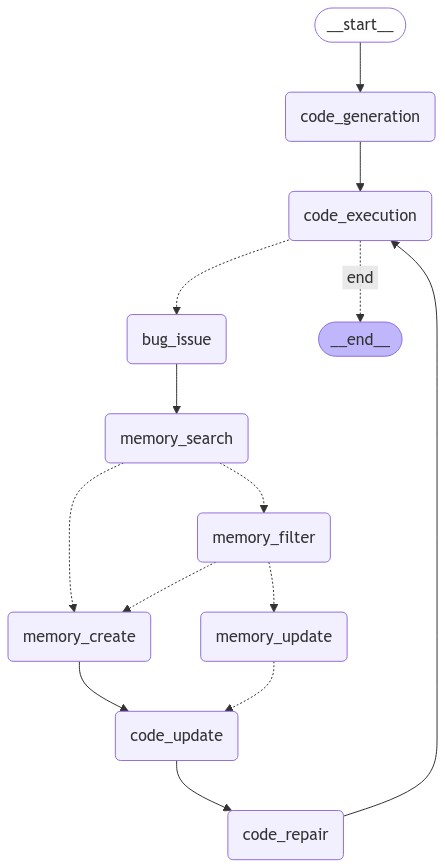}
\caption{Code Generation and Bug Fixing Flowchart Based on LangGraph.} \label{fig3}
\end{figure}

\begin{itemize}
    \item \textbf{Start State (start)}: The starting point of the entire process, marking the beginning of the LangGraph system's task execution.
    \item \textbf{Code Generation (code\_generation)}: After the start node, the system first performs code generation. Based on the prompt words described by the user, the GLM-4-Flash large model generates the corresponding Python code.
    \item \textbf{Code Execution (code\_execution)}: The generated Python code enters the execution phase, where the system automatically runs the generated code for verification. If the code executes successfully, the process directly reaches the end point (end), indicating that the code generation and execution process has no issues.
    \item \textbf{Report Bug Issue (bug\_issue)}: If a bug occurs during the execution of the Python code, the process enters the bug handling phase. The GLM-4-Flash large model generates an error report for the Python function that caused the bug.
    \item \textbf{Memory Search (memory\_search)}: When a bug occurs, the system performs a memory search. It looks for information related to the current bug in the bug knowledge base constructed by ChromaDB.
    \item \textbf{Memory Filter (memory\_filter)}: After searching for memory information, the system filters the search results to select the most relevant and most likely information to solve the current bug, ensuring that the system focuses only on the most relevant and useful information.
    \item \textbf{Memory Create (memory\_create)}: Based on the new bug report, the GLM-4-Flash large model generates relevant records and stores them in the vector database.
    \item \textbf{Memory Update (memory\_update)}: Based on the current bug report and the information from previous memory records, the GLM-4-Flash large model generates an updated bug report summary and updates the memory records in the vector database.
    \item \textbf{Code Update (code\_update)}: The buggy function code and bug issue description are sent to the GLM-4-Flash large model to regenerate the Python function code after bug fixing.
    \item \textbf{Code Repair (code\_repair)}: In LangGraph, the Python function is repaired and updated to the code after bug fixing. The code is executed again. If the code after bug fixing executes successfully, it indicates that the LangGraph process is complete and the new code has no issues; if the code still has bugs, the above process is iterated until the entire code bug fixing process is completed.
\end{itemize}

\subsection{case study}
This paper provides two examples, which respectively demonstrate the code generation and bug fixing for calculating the area of a triangle and dividing two numbers.

As shown in Figure 4, the process of using Python to calculate the area of a triangle is illustrated. The function is used to compute the area of a triangle.
\begin{figure}
\centering
\includegraphics[width=1\textwidth]{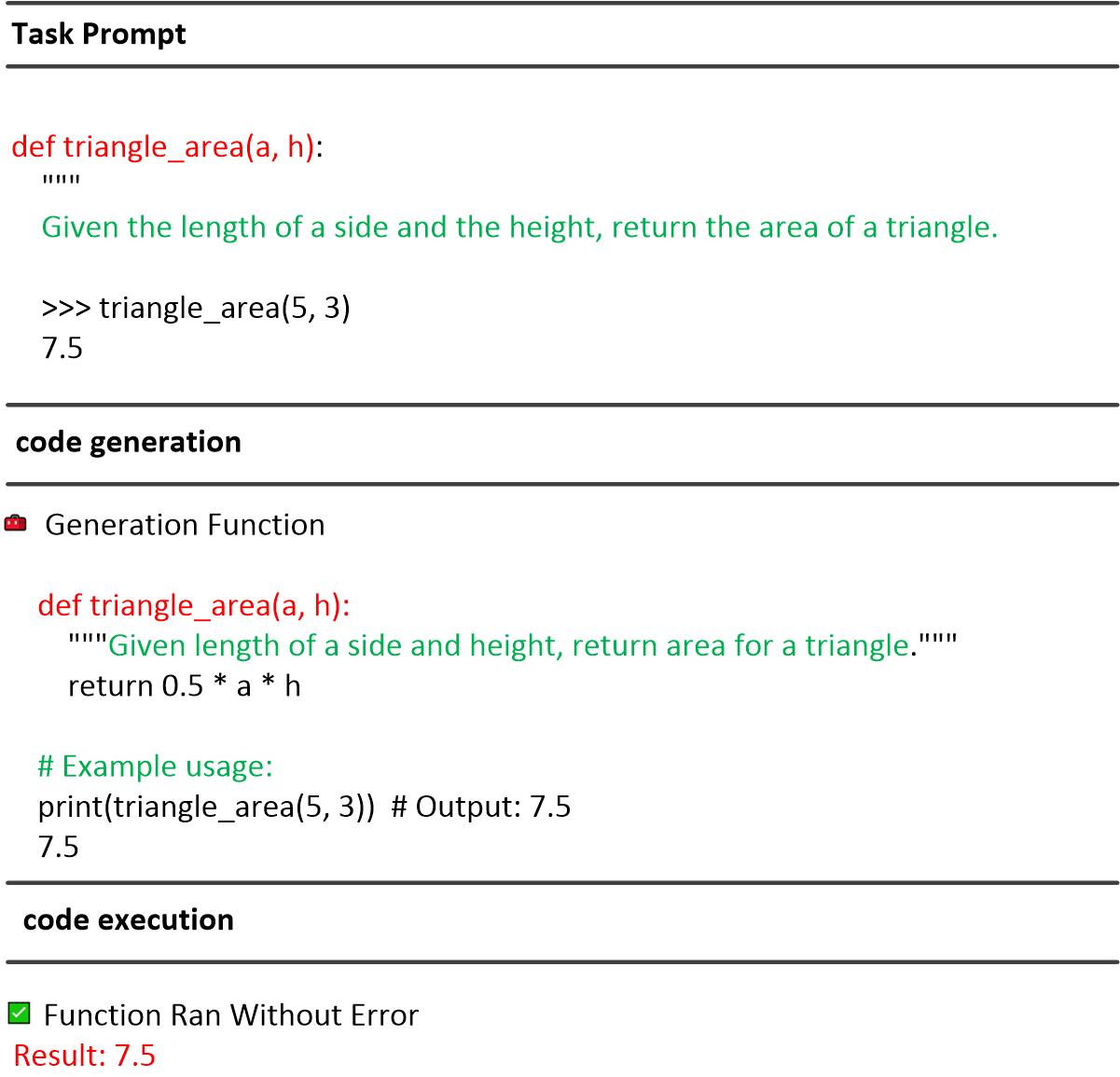}
\caption{The code implementation process for calculating the area of a triangle.} \label{fig4}
\end{figure}

\begin{itemize}
    \item \textbf{Task Description}: A clear task description is defined, stating that the main purpose of the function is to calculate the area of a triangle based on its given side length and height.
    \item \textbf{Code Implementation}: The function code for calculating the area of a triangle is generated using the GLM-4-Flash large model.
    \item \textbf{Code Execution}: The triangle area calculation function is verified. During execution, no errors were encountered, and the function returned a result of 7.5, which matches the expected result, proving the correctness of the function.
\end{itemize}
The above steps outline the process of calculating the area of a triangle using a Python function, including task description, code implementation, and code verification. Throughout the entire execution process, no bugs occurred, and the LangGraph code ran smoothly until completion, ultimately yielding accurate results.
 
As shown in Figure 5, the entire process of the Python programming task for dividing two numbers is illustrated, including task prompts, code generation, code execution, error reporting, code updating, and the execution results of the repaired code.
\begin{figure}
\centering
\includegraphics[width=0.8\textwidth]{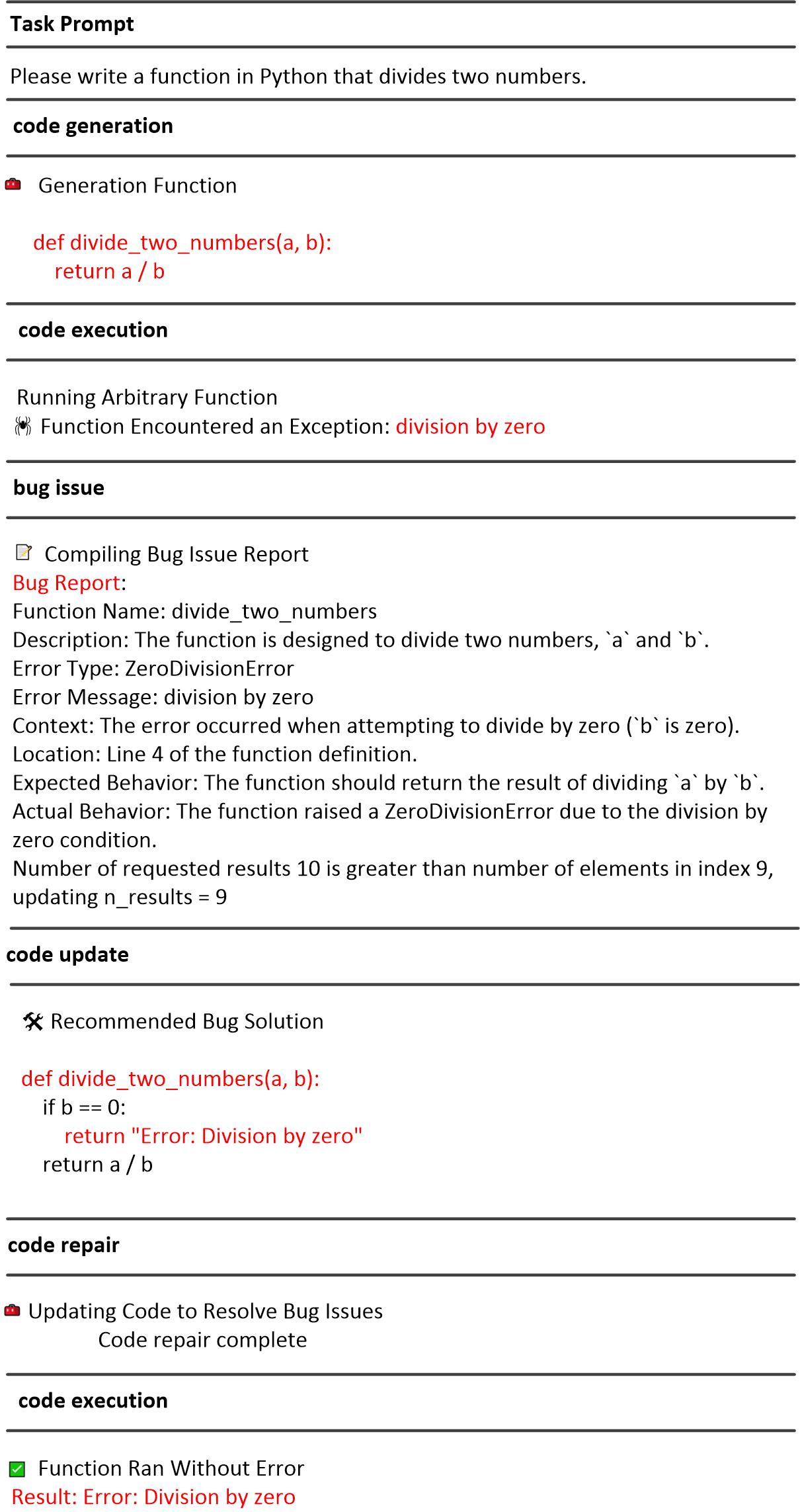}
\caption{The implementation process of a function for dividing two numbers
.} \label{fig5}
\end{figure} 

\begin{itemize}
    \item \textbf{Task Prompt}: Please write a Python function that can divide two numbers.
    \item \textbf{Code Generation}: The "divide\_two\_numbers" function is generated through the GLM-4-Flash large model, accepting two parameters and returning the division result.
    \item \textbf{Code Execution}: An exception occurs during function execution when the divisor is zero.
    \item \textbf{BUG Report}: Detailed error information is provided, including the function name, error type, error message, context of the error, location, and expected versus actual behavior. Similar bug records are retrieved from the vector database to obtain the number of request results and indices. The vector-based memory system, implemented using ChromaDB, enables efficient storage and supports semantic search for bug patterns, maintaining contextual relationships between bugs, and supporting pattern-based learning, as well as processes such as memory search, memory filtering, memory creation, and memory updating.
    \item \textbf{Code Update}: The GLM-4-Flash large model provides a solution by adding a conditional check to return a specific error message if the divisor is 0.
    \item \textbf{Code Repair}: The function code status is updated based on LangGraph, adding handling for the case where the divisor is 0.
    \item \textbf{Execution After Code Repair}: This time, the function runs without errors and returns the result "Error: Division by zero".
\end{itemize}

The overall process of generating the code for dividing two numbers demonstrates how a system based on large models and LangGraph can avoid division-by-zero errors in the division function by adding error handling.

\section{Discussion}

\subsection{  Large Language Model}
Regarding code generation and bug fixing, enhancing the reasoning capabilities of large language models is crucial. This enables the models to accurately identify information related to bugs in the code and precisely locate the specific positions of the bugs. With the powerful code reasoning capabilities of large models, bugs can be effectively fixed by targeting the most relevant parts of the code.

\subsection{Agent}
From the perspective of an intelligent agent, the proxy needs to focus on multiple error locations in the bug stack. By constructing a bug vector knowledge base, finer-grained localization of bugs can be achieved. Additionally, establishing a function verification mechanism or introducing third-party testing and verification functions can further validate the logical relationships of the functions. These measures help prevent the recurrence of bugs, thereby preventing code generation failures.

\section{Conclusion}
This study proposes a code self-repair system based on LangGraph and large models. The system achieves precise control of the process through LangGraph, maintaining a unified state object during runtime and updating the state according to node flow. The system constructs four major proxy components for code generation, execution, repair, and updating, and improves the accuracy of bug fixing through a structured bug vector knowledge base. Through iterative verification, the generated code can run smoothly and meet application requirements.

\section{Acknowledgement}
The authors express their gratitude to GLM-4-Flash from Zhipu AI Technology Co., Ltd. in Beijing, China, for their assistance with large language models. 
 
 
 

\end{document}